\documentclass[aps,pre,twocolumn,showpacs]{revtex4-1}
\usepackage{epsfig}
\usepackage{bm}

\newcommand{\be}{\begin{eqnarray}}
\newcommand{\ee}{\end{eqnarray}}
\newcommand{\bmat}{\left(\begin{array}}
\newcommand{\emat}{\end{array}\right)}
\newcommand{\no}{\nonumber}

\begin{document}
\title{Transitionless quantum driving for spin systems}
\author{Kazutaka Takahashi}
\affiliation{Department of Physics, Tokyo Institute of Technology, 
Tokyo 152-8551, Japan}

\date{\today}

\begin{abstract}
We apply the method of transitionless quantum driving 
for time-dependent quantum systems to spin systems. 
For a given Hamiltonian, the driving Hamiltonian is constructed 
so that the adiabatic states of the original system 
obey the Schr\"odinger equation. 
For several typical systems such as the $XY$ spin chain
and the Lipkin-Meshkov-Glick model, 
the driving Hamiltonian is constructed explicitly.
We discuss possible interesting situations 
when the driving Hamiltonian becomes time independent and 
when the driving Hamiltonian is equivalent to the original one.
For many-body systems, 
a crucial problem occurs at the quantum phase transition point where 
the energy gap between the ground and first excited states becomes zero. 
We discuss how the defect can be circumvented in the present method. 
\end{abstract}
\pacs{
05.30.-d, %Quantum statistical mechanics
03.67.Ac, %Quantum algorithms, protocols, and simulations 
75.10.Pq %Spin chain models 
}
\maketitle

%%%%%%%%%%%%%%%%%%%%%%%%%%%%%%%%%%%%%%%%%%%%%%%%%%%%%%%%%%%%%%%%%%%%%%%%%%%%
%%%%%%%%%%%%%%%%%%%%%%%%%%%%%%%%%%%%%%%%%%%%%%%%%%%%%%%%%%%%%%%%%%%%%%%%%%%%
\section{Introduction}

Understanding the dynamics of a quantum system is 
a fundamental problem in quantum mechanics and 
is important for practical applications. 
With recent advances in experimental techniques, 
we need to control the dynamics in a high-precision way~\cite{WPW, WR}.
Examples are seen in 
Bose-Einstein condensates in optical lattices~\cite{KWW}, 
quantum computations~\cite{FGGLLP} and so on.
These advances motivate us to design the optimal Hamiltonian 
rather than solving the problem under a given Hamiltonian.

Theoretically, several acceleration methods have been discussed recently: 
assisted adiabatic passage~\cite{DR1, DR2}, 
quantum brachistochrone~\cite{CHKO1, CHKO2}, 
fast-forward method~\cite{MN1, MN2, MN3}, 
transitionless quantum driving (TQD)~\cite{Berry}, 
quantum adiabatic brachistochrone~\cite{RKHLZ} 
and Lewis-Riesenfeld invariant-based 
engineering~\cite{LR, CRSCGM}.
In these methods, we discuss how fast one can reach 
a desired state starting from an initial one.
We construct a time-dependent Hamiltonian so that 
the time evolution is achieved efficiently.

These methods have been applied to systems with few degrees of freedom 
such as the Landau-Zener two-level system and 
a particle in a harmonic oscillator potential.
In the present paper, 
we treat systems with many degrees of freedom
by using the method of the TQD~\cite{Berry}, 
or, equivalently, the assisted adiabatic passage~\cite{DR1, DR2}.
For a given Hamiltonian, we construct the driving Hamiltonian 
so that the adiabatic states of the original Hamiltonian obey 
the Schr\"odinger equation.
The advantage of using this method is in its simplicity.
The generalization to many-body systems is a straightforward task.
The method has been tested experimentally in Bose-Einstein condensates 
in optical lattices and is shown to be robust against 
control parameter variations~\cite{Betal}.

Although the essential point of the method can be seen in small systems, 
we want to discuss the potential usefulness of the method 
when it is applied to many-body systems.
We can find a lot of interesting phenomena in such systems.
The effects of quantum fluctuations play important roles there.
Since the method of the TQD manipulates the same effect, 
we expect that the present analysis reveals the nature 
of the quantum effects from a different aspect.
As there are many interesting phenomena and 
a lot of techniques have been established, 
it is worth applying the present method to spin systems.

Practically, the method will be most useful when 
the final state is a nontrivial unknown one.
Starting from the initial trivial state, 
we evolve the system to the nontrivial state 
by using the quantum fluctuations. 
Such an idea is known as the quantum adiabatic calculations or,  
more generally, the quantum annealing~\cite{FGSSD, KN, FGGS}.
The major problem of this method is that 
infinite times are required to reach the final state.
By using the method of the TQD, we expect that such a problem is improved. 
We can also find what kind of quantum fluctuations we should use 
for the time evolution.

When we consider many-body systems, 
one of the most interesting phenomena that cannot be seen in few-body ones
is the quantum phase transition.
It is known that the adiabatic approximation fails 
at the phase transition point 
where the energy gap between the first and ground states becomes zero. 
This problem is not improved in the present method
since the driving Hamiltonian is divergent at that point, 
which can easily be understood from the general formula~\cite{Berry}.
However, we discuss that this is not a disaster and we can possibly 
circumvent the problem by using quantum fluctuations.

The paper is organized as follows.
In the next section, we review the method briefly 
and discuss possible extensions.
Then, we discuss 
a two-level system in Sec.~\ref{sec:tl}, 
a two-spin system in Sec.~\ref{sec:ts},
the one-dimensional $XY$ model in Sec.~\ref{sec:xy},
and the Lipkin-Meshkov-Glick model in Sec.~\ref{sec:LMG}.
Section~\ref{sec:summary} is devoted to summary.

%%%%%%%%%%%%%%%%%%%%%%%%%%%%%%%%%%%%%%%%%%%%%%%%%%%%%%%%%%%%%%%%%%%%%%%%%%%%
%%%%%%%%%%%%%%%%%%%%%%%%%%%%%%%%%%%%%%%%%%%%%%%%%%%%%%%%%%%%%%%%%%%%%%%%%%%%
\section{Transitionless quantum driving}
\label{sec:tqd}

%%%%%%%%%%%%%%%%%%%%%%%%%%%%%%%%%%%%%%%%%%%%%%%%%%%%%%%%%%%%%%%%%%%%%%%%%%%%
\subsection{Transitionless quantum driving}

We treat a time-dependent quantum Hamiltonian $\hat{H}_0(t)$.
To obtain the corresponding state under some initial condition, 
we solve the Schr\"odinger equation, 
which is usually a formidable task especially for many-body systems.
The analysis becomes considerably easier 
when we consider the slow evolution of the Hamiltonian.
In that case, the instantaneous eigenenergies and eigenstates 
defined as 
\be
 \hat{H}_0(t)|n(t)\rangle = E_n(t)|n(t)\rangle
 \label{inst}
\ee
determine the state of the system.
Here, $n$ is the index denoting each state.
Then, if we start the time evolution 
from one of the eigenstates $|n(t_0)\rangle$ at $t=t_0$, 
the adiabatic state is given by 
\be
 |\psi_n(t)\rangle &=& \exp\left(-i\int_{t_0}^t dt' E_n(t')\right. \no\\
 && \left.
 -\int_{t_0}^t dt' \langle n(t')|\frac{d}{dt'}|n(t')\rangle
 \right)|n(t)\rangle. 
 \label{ad}
\ee
The first term in the exponential represents 
the counterpart of the time-evolution factor in the stationary states 
and the second one is a phase factor that 
generates the geometric Berry phase~\cite{Berry2}.
The adiabatic state is a good approximation as the solution of 
the Schr\"odinger equation 
when the following adiabatic condition is satisfied: 
\be
 \frac{|\langle m(t)|\frac{d\hat{H}_0(t)}{dt}|n(t)\rangle|}
 {(E_m(t)-E_n(t))^2} \ll 1,
 \label{adcond}
\ee
where $m$ and $n$ denote different energy levels.

Berry found a formula of the driving Hamiltonian 
such that transitions to other states do not occur~\cite{Berry}.
This means that the adiabatic state becomes the exact solution of 
the Sch\"odinger equation 
\be
 i\frac{d}{dt}|\psi_n(t)\rangle
 = \hat{H}(t)|\psi_n(t)\rangle.
\ee
The new Hamiltonian $\hat{H}(t)$ is different from 
the original one $\hat{H}_0(t)$.
Applying the time derivative operator to the adiabatic state (\ref{ad}),
we can easily obtain $\hat{H}(t)=\hat{H}_0(t)+\hat{H}_1(t)$ with 
\be
 \hat{H}_1(t) &=&
 \sum_{m}\left(1-|m(t)\rangle\langle m(t)|\right)
 i\frac{d}{dt}|m(t)\rangle\langle m(t)|.
 \label{H1}
\ee
We note that the time-derivative operator acts only on 
the state just behind the operator.
If we consider the time-evolution under 
this new Hamiltonian $\hat{H}(t)$ 
starting from the initial state $|n(t_0)\rangle$, the state remains 
the same eigenstate of the original Hamiltonian $\hat{H}_0(t)$.
We can consider the time evolution in an arbitrary speed, 
which accomplishes our purpose to go beyond the adiabatic evolution.

As we mentioned in the introduction,
the driver Hamiltonian diverges at the level crossing point.
This can be understood from the alternative expression of 
the formula (\ref{H1}) as 
\be
 \hat{H}_1(t) =
 i\sum_{l,m\, (l\ne m)} |l(t)\rangle
 \frac{\langle l(t)|\frac{d\hat{H}_0(t)}{dt}|m(t)\rangle}{E_m(t)-E_l(t)}
 \langle m(t)|.
 \label{H1e}
\ee
This problem is crucial when we apply the method to the many-body systems 
where the level crossing occurs constantly. 
Even if we restrict ourselves to 
the time evolution of the ground state, 
the problem occurs at the quantum phase transition point 
where the energy gap between the ground and first excited 
states goes to zero.

The problem does not occur when the matrix element 
in Eq.~(\ref{H1e}) goes to zero.
Then, the driving Hamiltonian is equal to the original Hamiltonian: 
$\hat{H}(t)=\hat{H}_0(t)$.
The adiabatic state
becomes the exact solution of the Schr\"odinger equation 
and the time evolution can be done in an arbitrary speed.
This ``fixed-point'' condition means that 
the eigenstates $|n(t)\rangle$ of 
the original Hamiltonian $\hat{H}_0(t)$ are also the eigenstates of 
the time derivative of $\hat{H}_0(t)$.
That is, their operators commute with each other as follows: 
\be
 \left[\hat{H}_0(t), \frac{d}{dt}\hat{H}_0(t)\right] = 0.
 \label{commute}
\ee
We can also state this condition from Eq.~(\ref{H1}) that 
the time derivative of the eigenstate $|n(t)\rangle$
is proportional to $|n(t)\rangle$.

We note that this condition does not always give 
an interesting situation.
Equation (\ref{commute}) is satisfied 
when the time dependence of the Hamiltonian 
is written as $\hat{H}_0(t)= f(t)\hat{H}_0(t_0)$ where 
$f(t)$ is an arbitrary function of time with 
the initial condition $f(t_0)=1$.
In this case, 
the corresponding system is reduced to a static one 
by the time reparametrization $dt'=f(t)dt$. 
In the following, we find in some cases that 
the fixed-point condition gives a nontrivial result.

%%%%%%%%%%%%%%%%%%%%%%%%%%%%%%%%%%%%%%%%%%%%%%%%%%%%%%%%%%%%%%%%%%%%%%%%%%%%
\subsection{Extensions}

We discuss possible extensions of the Berry's method  
with the application to many-body systems in mind.
The original formula (\ref{H1}) by Berry was derived 
under the condition that there are no degeneracies of the eigenstates.
We used the orthonormal relation 
$\langle m(t)|n(t)\rangle=\delta_{mn}$ to derive the formula.
It is possible to extend the formula to systems with degenerate states.
The adiabatic state for such systems was discussed in \cite{WZ}.
We specify the instantaneous eigenstates as  
\be
 \hat{H}_0(t)|n,\mu,t\rangle = E_n(t)|n,\mu,t\rangle.
\ee
Each eigenstate is labeled by two indices $n$ and $\mu$.
It has an eigenenergy $E_n$ and 
the number of values of $\mu$ represents the degeneracy.
For the degenerate states we can choose 
the appropriate basis such that the orthonormal relation 
\be
 \langle n,\mu,t|n',\mu',t\rangle = \delta_{nn'}\delta_{\mu\mu'}
\ee
is satisfied.
Then, the adiabatic state is written as 
\be
 |\psi_n(t)\rangle = \sum_{\mu}
 \exp\left(-i\int_{t_0}^t dt' E_n(t')\right)
 c_{\mu}^{(n)}(t)|n,\mu,t\rangle. \no\\
\ee
The coefficients $c_{\mu}^{(n)}$ are determined below.
Inserting this expression to the Schr\"odinger equation and 
using the adiabatic approximation, we obtain 
\be
 \frac{d}{dt}c_{\mu}^{(n)}(t) = -\sum_{\mu'}
 \langle n,\mu,t|\frac{d}{dt} |n,\mu',t\rangle c_{\mu'}^{(n)}(t).
\ee
This differential equation is solved as 
\be
 c_{\mu}^{(n)}(t) = \sum_{\mu'}U_{\mu\mu'}^{(n)}(t)c_{\mu'}^{(n)}(t_0), 
\ee
where $U_{\mu\mu'}^{(n)}(t)$ is a matrix element of the matrix 
\be
 U^{(n)}(t) = {\rm T}\exp\left(
 -i\int_{t_0}^t dt' A^{(n)}(t')\right).
\ee
The symbol ${\rm T}$ represents the time-ordered product and 
the non-Abelian gauge field matrix $A^{(n)}$ 
has its element as 
\be
 iA_{\mu\mu'}^{(n)}(t)=
 \langle n,\mu,t|\frac{d}{dt} |n,\mu',t\rangle.
\ee
In this adiabatic state, 
it is straightforward to apply the method of the TQD.
After a similar calculation as the original case, we obtain 
\be
 \hat{H}_1(t)=\sum_{n\mu}
 \left(1-\sum_{\nu}|n,\nu,t\rangle\langle n,\nu,t|\right)
 i\frac{d}{dt}|n,\mu,t\rangle\langle n,\mu,t|. \no\\
 \label{H1deg}
\ee
We note that the formula (\ref{H1deg}) is applied when 
the degeneracies are maintained throughout the time evolution.
The method does not work well
when the level crossing occurs.

Next, we consider a more useful extension for many-body systems.
The original formula (\ref{H1}) is applied for any instantaneous energy levels.
However, for many-body systems, we usually are interested 
in the ground state only.
Then, if operators give nothing when they are applied to the ground state, 
they can be neglected from the beginning.
If we only consider some specific energy level $n$,
we can use the state-dependent driver Hamiltonian 
\be
 \hat{H}_1^{(n)}(t) = \left(1-|n(t)\rangle\langle n(t)|\right)
 i\frac{d}{dt}|n(t)\rangle\langle n(t)| + ({\rm H.c.}), \no\\
 \label{H1n}
\ee
where $({\rm H.c.})$ denotes the Hermite conjugate of the first term.
Thus, we have an arbitrariness in choosing the driver Hamiltonian.
This is a small finding but is very useful for practical applications.

%%%%%%%%%%%%%%%%%%%%%%%%%%%%%%%%%%%%%%%%%%%%%%%%%%%%%%%%%%%%%%%%%%%%%%%%%%%%
%%%%%%%%%%%%%%%%%%%%%%%%%%%%%%%%%%%%%%%%%%%%%%%%%%%%%%%%%%%%%%%%%%%%%%%%%%%%
\section{Two-level system}
\label{sec:tl}

We can see how the method works well in a simple two level Hamiltonian 
\be
 \hat{H}_0(t) = \bm{h}^{(0)}(t)\cdot\hat{\bm{\sigma}}, 
\ee
where $\bm{h}^{(0)}(t)$ is a magnetic field vector in three dimensional space 
and $\hat{\bm{\sigma}}$ are Pauli matrices.
This case was considered in \cite{Berry}. 
Applying the general formula to this Hamiltonian, we obtain 
the driving Hamiltonian $\hat{H}(t)=\hat{H}_0(t)+\hat{H}_1(t)$ with 
\be
 \hat{H}_1(t) = \frac{1}{2|\bm{h}^{(0)}(t)|^2}
 \left(\bm{h}^{(0)}(t)\times\dot{\bm{h}}^{(0)}(t)
 \right)\cdot\hat{\bm{\sigma}},
 \label{h1}
\ee
where $\dot{\bm{h}}^{(0)}(t)$ represents the time derivative of 
$\bm{h}^{(0)}(t)$.
Thus, the magnetic field to apply is given by 
$\bm{h}(t)=\bm{h}^{(0)}(t)
+(\bm{h}^{(0)}(t)\times\dot{\bm{h}}^{(0)}(t))/2|\bm{h}^{(0)}(t)|^2$.

This simple example tells us an important fact about 
the nature of quantum fluctuations.
Even if we confine the magnetic field in a plane, e.g., 
$\bm{h}^{(0)}(t)=(h_1(t), h_2(t), 0)$, 
the driver Hamiltonian gives a field in the $z$ direction.
The quantum fluctuation effects unavoidably produce 
all kinds of operators which act on states in the Hilbert space.
In other words, we need a complex matrix to represent
the driving Hamiltonian.

As a simple example, we consider an oscillating field 
\be
 \bm{h}^{(0)}(t) = \bmat{c} h_0\cos\omega t \\ 
 h_0\sin \omega t \\ h_3 \emat. \label{h0o}
\ee
Then, the driving field has the same form as the original one as
\be
 \bm{h}(t) = \bmat{c} \tilde{h}_0\cos\omega t \\ 
 \tilde{h}_0\sin \omega t \\ \tilde{h}_3 \emat, \label{ho}
\ee
where 
\be
 & & \tilde{h}_0 = h_0\left(1-\frac{\omega h_3}{2(h_0^2+h_3^2)}\right), \\
 & & \tilde{h}_3 = h_3+\frac{\omega h_0^2}{2(h_0^2+h_3^2)}.
\ee
Thus, by considering the time evolution under the field (\ref{ho}), 
we can obtain the adiabatic state of the original field (\ref{h0o}).

It is interesting to consider some special cases.
The first one is when the driving Hamiltonian becomes time independent.
The condition $\dot{\bm{h}}=0$ gives 
\be
 \dot{\bm{h}}^{(0)}(t)
 +\frac{1}{2h_0^2(t)}\bm{h}^{(0)}(t)\times\ddot{\bm{h}}^{(0)}(t) = 0.
\ee
We see that $\dot{\bm{h}}^{(0)}(t)$ is perpendicular to 
$\bm{h}^{(0)}(t)$ and $\ddot{\bm{h}}^{(0)}(t)$.
In the case of the oscillating field (\ref{h0o}),
the condition $\tilde{h}_0=0$ gives 
\be
 2(h_0^2+h_3^2)=\omega h_3. \label{h001}
\ee
Then, the driving Hamiltonian has the static field 
\be
 \bm{h}(t) = \bmat{c} 0 \\ 0 \\ \frac{\omega}{2} \emat.
\ee
Driving the system by the static Hamiltonian $\hat{H}$,  
we can find the adiabatic state of the original Hamiltonian.
We note that 
we must choose one of the eigenstates of $\hat{H}_0(t_0)$
as the initial condition which is not the eigenstate of $\hat{H}$.

The second case to be examined is 
when the fixed-point condition $\bm{h}(t)=\bm{h}^{(0)}(t)$ is satisfied.
This equation is easily solved and 
we obtain $\bm{h}^{(0)}(t)=f(t)\bm{h}^{(0)}(t_0)$ where 
$f(t)$ is an arbitrary function of time with 
the initial condition $f(t_0)=1$.
This means that the magnetic field points to the same direction 
throughout the time evolution.
In this case, the corresponding system is reduced to a static one 
as we discussed in the previous section and 
the fixed-point condition does not play an important role.
To find a nontrivial situation, 
we need to consider more complicated systems.

%%%%%%%%%%%%%%%%%%%%%%%%%%%%%%%%%%%%%%%%%%%%%%%%%%%%%%%%%%%%%%%%%%%%%%%%%%%%
%%%%%%%%%%%%%%%%%%%%%%%%%%%%%%%%%%%%%%%%%%%%%%%%%%%%%%%%%%%%%%%%%%%%%%%%%%%%
\section{Two-spin system}
\label{sec:ts}

As a preliminary calculation to many-body systems, 
we consider a two-spin system.
This is the simplest many-body system and we can learn 
typical problems which occur in general many-body systems.
Furthermore, the model is directly applied to two-qubit systems.

The Hamiltonian is given by
\be
 \hat{H}_0(t)
 = J_x(t)\sigma_1^x\sigma_2^x
 +J_y(t)\sigma_1^y\sigma_2^y
 +h(t)(\sigma_1^z+\sigma_2^z), \label{2spin}
\ee
where $\sigma_1^{x,y,z}$ and $\sigma_2^{x,y,z}$ represent 
Pauli matrices for two kinds of spins.
We write this Hamiltonian by using the $z$ basis 
$|m_1,m_2\rangle=|m_1\rangle_1|m_2\rangle_2$ 
where $\sigma_1^z|m\rangle_1=m|m\rangle_1$ and 
$\sigma_2^z|m\rangle=m|m\rangle_2$ with $m=\pm 1$.
Arranging the basis states as 
$|+1,+1\rangle$, $|-1,-1\rangle$, 
$|+1,-1\rangle$ and $|-1,+1\rangle$,
we obtain a two block form 
\be
 \hat{H}_0 = \bmat{cccc}
 2h & J_x-J_y & 0 & 0 \\
 J_x-J_y & -2h & 0 & 0 \\
 0 & 0 & 0 & J_x+J_y \\
 0 & 0 & J_x+J_y & 0 \emat.  \label{mat}
\ee
The matrix in each block is understood as the Hamiltonian of 
a two-level system and we can apply the formula in Eq.~(\ref{h1}). 
The second block represents a system with 
the magnetic field only in one direction and 
no further transformation is required.
Applying the formula (\ref{h1}) to the first block, 
we obtain the driver Hamiltonian in the space of the first block 
\be
 \hat{H}_1^{(1)}(t) = \frac{h(\dot{J}_x-\dot{J}_y)-\dot{h}(J_x-J_y)}
 {4h^2+(J_x-J_y)^2}\bmat{cc} 0 & -i \\ i & 0 \emat.
\ee
This representation is transformed to that in 
the original full space as 
\be
 \hat{H}_1(t)=\frac{1}{2}\frac{h(\dot{J}_x-\dot{J}_y)-\dot{h}(J_x-J_y)}
 {4h^2+(J_x-J_y)^2}\left(
 \sigma_1^x\sigma_2^y+\sigma_1^y\sigma_2^x\right). \no\\
\ee

To obtain a more useful form, we consider the unitary rotation 
\be
 \tilde{H}(t)=\hat{U}(t)
 \left(\hat{H}_0(t)+\hat{H}_1(t)\right)\hat{U}^\dag(t), 
\ee
where
\be
 \hat{U}(t)=\exp\left[-\frac{i}{2}\theta(t)(\sigma_1^z+\sigma_2^z)\right].
\ee
The rotation is performed around the $z$ axis and 
the angle $\theta$ is determined so that the transformed 
Hamiltonian has the same form as the original one.
The condition is given by 
\be
 \tan 2\theta = \frac{1}{J_x-J_y}
 \frac{\dot{h}(J_x-J_y)-h(\dot{J}_x-\dot{J}_y)}{4h^2+(J_x-J_y)^2}. 
\ee

Since the unitary rotation has a time dependence, 
the transformed state 
\be
 |\tilde{\psi}(t)\rangle = \hat{U}(t)|\psi(t)\rangle
\ee
obeys the Schr\"odinger equation with the Hamiltonian 
\be
 \hat{H}(t) &=& \tilde{H}(t)+\frac{\dot{\theta}(t)}{2}(\sigma_1^z+\sigma_2^z) \\
 &=& \tilde{J}_x(t)\sigma_1^x\sigma_2^x
 +\tilde{J}_y(t)\sigma_1^y\sigma_2^y
 +\tilde{h}(t)(\sigma_1^z+\sigma_2^z),
\ee
 where 
\be
 & & \tilde{J}_x(t)+\tilde{J}_y(t)=J_x(t)+J_y(t), \\
 & & \left(\tilde{J}_x(t)-\tilde{J}_y(t)\right)\cos 2\theta(t)=J_x(t)-J_y(t), \\
 & & \tilde{h}(t) = h(t)+\frac{\dot{\theta}(t)}{2}.
\ee
Thus, the time evolution of the adiabatic state 
of the original $XY$ Hamiltonian (\ref{2spin})
is described by the same Hamiltonian with different coupling constants.
If we consider the time evolution of a state 
for a given Hamiltonian $\hat{H}(t)$,
the state must be the adiabatic state of the Hamiltonian $\hat{H}_0(t)$.

We can also find in the present system 
the fixed point where $\hat{H}_1(t)$ becomes zero.
This condition is achieved by the magnetic field 
\be
 h(t)=c(J_x(t)-J_y(t)), 
\ee
where $c$ is a constant.
Although this condition is a trivial one 
in the first block of the matrix (\ref{mat}), 
it is not in the whole space of the Hamiltonian.
The system cannot be reduced to a static one  
by the time reparametrization.
%We expect that such a fixed point has a special meaning 
%for the dynamical problem, which we have not understood yet.

%%%%%%%%%%%%%%%%%%%%%%%%%%%%%%%%%%%%%%%%%%%%%%%%%%%%%%%%%%%%%%%%%%%%%%%%%%%%
%%%%%%%%%%%%%%%%%%%%%%%%%%%%%%%%%%%%%%%%%%%%%%%%%%%%%%%%%%%%%%%%%%%%%%%%%%%%
\section{One-dimensional anisotropic $\bm{XY}$ model}
\label{sec:xy}

As the simplest many-body system showing nontrivial ground-state properties, 
we study the one-dimensional anisotropic $XY$ model.
This model including the transverse-field Ising model is exactly solvable 
and has been used frequently to understand 
the quantum dynamics~\cite{Dziarmaga, KCH}.
We consider $N$ kinds of spins and write the Hamiltonian as 
\be
 \hat{H}_0(t)=-\sum_{j=1}^N
 \left(J_x(t)\sigma_j^x\sigma_{j+1}^x
 +J_y(t)\sigma_j^y\sigma_{j+1}^y
 -h(t)\sigma_j^z\right). \no\\
\ee
Taking $J_x=0$ or $J_y=0$, we can also study 
the transverse-field Ising model.
In the static case, this model has a quantum phase transition at 
$J_x+J_y=h$ where the energy gap between 
the ground and first excited states goes to zero.

This Hamiltonian is diagonalized by the method of fermionization.
We define the Jordan-Wigner transformation~\cite{JW, LSM}
\be
 & & \frac{1}{2}\left(\sigma_j^x+i\sigma_j^y\right) = 
 \exp\left(i\pi\sum_{k=1}^{j-1}a_k^\dag a_k\right)a_j^\dag, \\
 & & \sigma_j^z = 2a_j^\dag a_j -1.
\ee
Here, the operators $a_j$ obey the fermionic anti-commutation relations
\be
 & & a_ia_j^\dag+a_j^\dag a_i = \delta_{ij}, \\
 & & a_ia_j+a_ja_i = 0.
\ee
We also introduce the Fourier transformation 
\be
 a_j = \frac{1}{\sqrt{N}}\sum_q \tilde{a}_qe^{iqj}.
\ee
The momentum $q$ takes a value between $-\pi$ and $\pi$, and 
its discrete value depends on the parities of
$N$ and the number of fermions in each state. 
Since the value becomes continuous at the thermodynamic limit, 
we do not specify it explicitly here.
Applying the transformation to the Hamiltonian,
we obtain $\hat{H}_0(t)=\sum_q\hat{H}_0(q,t)$ where 
\be
 \hat{H}_0(q,t) &=& -2\left[(J_x+J_y)\cos q-h\right]
 \tilde{a}_q^\dag \tilde{a}_q \no\\
 & & -i(J_x-J_y)\sin q (\tilde{a}_q^\dag \tilde{a}_{-q}^\dag
 +\tilde{a}_q \tilde{a}_{-q})-h. \no\\
\ee
The Hamiltonian is block-diagonalized and 
each block is specified by the absolute value of $q$.
In each block, the state is specified explicitly by 
$|0\rangle$, $\tilde{a}_q^\dag \tilde{a}_{-q}^\dag|0\rangle$, 
$\tilde{a}_q^\dag|0\rangle$ and $\tilde{a}_{-q}^\dag|0\rangle$ where 
$|0\rangle$ is the vacuum of fermions defined by $\tilde{a}_q|0\rangle = 0$.
The Hamiltonian is given by 
$\hat{H}_0(q,t)+\hat{H}_0(-q,t)=\hat{H}_0^{(1)}(q,t)\otimes 1_2
+1_2\otimes[-2(J_x+J_y)\cos q]1_2$ where 
\begin{widetext}
\be
 \hat{H}_0^{(1)}(q,t)
 = \bmat{cc}
 -2h & 2i(J_x-J_y)\sin q \\
 -2i(J_x-J_y)\sin q & -4[(J_x+J_y)\cos q-h]-2h \emat.
\ee
%\end{widetext}
We have a two-level system again and the driving Hamiltonian is 
constructed in the same way as the previous calculations.
It is written as 
\be
 \hat{H}_1^{(1)}(q,t) = \frac{J_1(q,t)}{2}\sin q 
 \bmat{cc} 1 & 0 \\ 0 & 1 \emat,
\ee
 where 
%\begin{widetext}
\be
 J_1(q,t) = 
 \frac{[(J_x+J_y)\cos q-h](\dot{J}_x-\dot{J}_y)
 -(J_x-J_y)[(\dot{J_x}+\dot{J}_y)\cos q-\dot{h}]}
 {[(J_x+J_y)\cos q-h]^2+(J_x-J_y)^2\sin^2 q}. \no\\
\ee
\end{widetext}
In the full space of fermionic states,
the Hamiltonian is written in terms of fermion operators as  
$\hat{H}_1(t)=\sum_q\hat{H}_1(q,t)$ with 
\be
 \hat{H}_1(q,t)
 =\frac{J_1(q,t)}{4}\sin q \left(\tilde{a}_q^\dag 
 \tilde{a}_{-q}^\dag+\tilde{a}_{-q}\tilde{a}_q\right).
 \label{h1fermi}
\ee

The present purpose is accomplished 
if we represent this term by spin operators.
However, the representation (\ref{h1fermi}) has 
a $q$-dependent coupling 
which gives a nonlocal hopping of fermions.
As a result, the corresponding spin representation has
nonlocal and many-body interacting terms.
Unfortunately, the method gives an unrealizable Hamiltonian.
Although this term is treated by the truncation approximation
in \cite{dCRZ} for the transverse-field Ising model, 
we need infinite numbers of operators at the thermodynamic limit,
which makes the analysis difficult.

This problem does not arise if we restrict ourselves to a specific state.
In most of applications, we are interested in the ground state of 
the Hamiltonian.
The driving Hamiltonian is tuned so that 
the low-lying states are controlled properly.
In the present case, the low-lying states are denoted by 
excitations of the modes at small $q$.
The momentum $q$ in the coupling $J_1(q,t)$ is neglected and 
we have the driving Hamiltonian for the low-lying states 
\be
 \hat{H}_1(t)\sim \frac{J_1(t)}{4}\sum_q\sin q 
 \left(\tilde{a}_q^\dag \tilde{a}_{-q}^\dag+\tilde{a}_{-q}\tilde{a}_q\right),
\ee
 where 
\be
 J_1(t) &=& 
 \frac{(J_x+J_y-h)(\dot{J}_x-\dot{J}_y)}{(J_x+J_y-h)^2} \no\\
 && -\frac{(J_x-J_y)(\dot{J_x}+\dot{J}_y-\dot{h})}
 {(J_x+J_y-h)^2}.
\ee
In this case, it is easy to go back to the spin representation and 
we obtain 
\be
 \hat{H}_1(t) 
 &=& -i\frac{J_1(t)}{4}\sum_{j=1}^N 
 \left(a_j^\dag a_{j+1}^\dag+a_{j}a_{j+1}\right) \\
 &=& \frac{J_1(t)}{8}\sum_{j=1}^N 
 \left(\sigma_j^x\sigma_{j+1}^y+\sigma_{j}^y\sigma_{j+1}^x\right).
\ee
If we consider the unitary rotation as was done in the two-spin system, 
we can have as the total driving Hamiltonian 
\be
 \hat{H}(t)=
 -\sum_{j=1}^N
 \left(\tilde{J}_x(t)\sigma_j^x\sigma_{j+1}^x
 +\tilde{J}_y(t)\sigma_j^y\sigma_{j+1}^y
 -\tilde{h}(t)\sigma_j^z\right), \no\\
\ee
where $\tilde{J}_x(t)$, $\tilde{J}_y(t)$ and $\tilde{h}(t)$ 
are determined properly.

We note that $J_1(t)$ is divergent at the phase transition point $J_x+J_y=h$ 
as we can understand from the general formula (\ref{H1e}).
This problem can be circumvented if we choose 
the fixed-point condition $J_1(t)=0$.
It gives the driving protocol 
\be
 h(t)=J_x(t)+J_y(t)+c(J_x(t)-J_y(t)), 
\ee
where $c$ is an arbitrary constant.
In this protocol, the phase transition takes place at 
the isotropic point $J_x(t)=J_y(t)$.
In the isotropic $XY$ model, 
it is known that the singularity at the point $J_x=J_y=h$ is weak.
It is interesting to see that 
the fixed-point condition chooses such a point properly 
to go beyond the phase boundary.

%%%%%%%%%%%%%%%%%%%%%%%%%%%%%%%%%%%%%%%%%%%%%%%%%%%%%%%%%%%%%%%%%%%%%%%%%%%%
%%%%%%%%%%%%%%%%%%%%%%%%%%%%%%%%%%%%%%%%%%%%%%%%%%%%%%%%%%%%%%%%%%%%%%%%%%%%
\section{Lipkin-Meshkov-Glick model}
\label{sec:LMG}

%%%%%%%%%%%%%%%%%%%%%%%%%%%%%%%%%%%%%%%%%%%%%%%%%%%%%%%%%%%%%%%%%%%%%%%%%%%%
\subsection{Analytic result}

Then, we move to the analysis of 
the $XY$ model with infinite-range interaction known as 
the Lipkin-Meshkov-Glick model~\cite{LMG}.
Using this model, we examine more closely 
how the fixed-point condition is effectively utilized.

The Hamiltonian is given by 
\be
 \hat{H}_0(t) = -\frac{2J_x(t)}{N}(\hat{S}^x)^2
 -\frac{2J_y(t)}{N}(\hat{S}^y)^2
 -2h(t)\hat{S}^z,
\ee
 where $\hat{\bm{S}}=(\hat{S}^x,\hat{S}^y,\hat{S}^z)$ is 
 the sum of the Pauli operator at each site
\be
 \hat{\bm{S}}=\frac{1}{2}\sum_{i=1}^N\bm{\sigma}_i.
\ee
To avoid unnecessary complications, we assume that 
time-dependent parameters $J_x(t)$, $J_y(t)$, and $h(t)$ are positive.

The Hamiltonian is expressed by the total spin, which means that 
the quantum number $\hat{\bm{S}}^2=S(S+1)$ is conserved.
The ground state belongs to the sector with the maximum spin $S=N/2$,
and by taking the thermodynamic limit $N\to\infty$
we can use the semiclassical method.
The ground state is determined classically and 
we parametrize the spin configuration as 
\be
 \bm{S}=\frac{N}{2}\bmat{c} 
 \sin\theta\cos\varphi \\ \sin\theta\sin\varphi \\ \cos\theta \emat.
\ee
This expression is inserted to the Hamiltonian and 
the minimum-energy condition gives the solution 
\be
 \begin{array}{ll}
 \theta = 0 & h\ge  J_{x},\ h\ge  J_{y}  \\
 \varphi=0,\ \cos\theta = \frac{h}{J_x} & h< J_x,\ J_x> J_y \\
 \varphi=\frac{\pi}{2},\ \cos\theta = \frac{h}{J_y} & h< J_y,\ J_x< J_y  
 \end{array}.
\ee
We see that the quantum phase transition occurs at the point
$h={\rm max}(J_x,J_y)$.
If the magnetic field is larger than this value,
all spins align in the $z$ direction 
with small quantum corrections discussed below.
For smaller fields, we have a symmetry-breaking state.

First, we consider the symmetric phase with 
$h\ge {\rm max}(J_x,J_y)$.
In this case, excitations from the ground state are treated semiclassically.
We use the Holstein-Primakoff transformation~\cite{HP}
\be
 & & \hat{S}^z = S-b^\dag b, \\
 & & \hat{S}^x+i\hat{S}^y = \sqrt{2S-b^\dag b}\,b, 
\ee
where $b$ denotes the bosonic operator satisfying $[b,b^\dag]=1$.
At the thermodynamic limit, 
the Hamiltonian is expanded with respect to $1/N$.
Up to zeroth order in $1/N$, we have 
$\hat{H}_0(t)\sim E_0(t)+\hat{h}_0(t)$ where 
$E_0(t)=-Nh(t)$ and 
\be
 \hat{h}_0(t) = -\frac{J_x}{2}(b+b^\dag)^2+\frac{J_x}{2}(b-b^\dag)^2
 +2h b^\dag b.
\ee
This Hamiltonian is diagonalized by the Bogoliubov transformation 
\be
 & & b=\alpha\cosh\frac{\Theta}{2}+\alpha^\dag\sinh\frac{\Theta}{2}, \\
 & & b^\dag=\alpha^\dag\cosh\frac{\Theta}{2}+\alpha\sinh\frac{\Theta}{2},
\ee
where the operator $\alpha$ denotes a different kind of bosons.
The parameter $\Theta$ is determined by the diagonalization condition 
\be
 \tanh\Theta=\frac{J_x-J_y}{2h-J_x-J_y}.
\ee
Then, we obtain the standard harmonic oscillator form 
\be
 & & \hat{h}_0(t) = 
 -E_0-h+\Omega \left(\alpha^\dag\alpha+\frac{1}{2}\right), \\
 & & \Omega = 2\sqrt{(h-J_x)(h-J_y)}.
\ee

Knowing the instantaneous eigenvalues and eigenstates, we can calculate 
the driver Hamiltonian.
After some calculations, we obtain 
\be
 \hat{H}_1(t) 
 = \frac{-ih_1(t)}{2}(b^2-(b^\dag)^2) 
 \sim \frac{h_1(t)}{N}(\hat{S}^x\hat{S}^y+\hat{S}^y\hat{S}^x), \no\\
\ee
where 
\be
 h_1(t)=\frac{
 (h-J_x)(\dot{h}-\dot{J}_y)-(h-J_y)(\dot{h}-\dot{J}_x)}
 {2(h-J_x)(h-J_y)}.
\ee
We note that this driver Hamiltonian works only for the ground state
as we discuss below.

The fixed-point condition is given by $h_1(t)=0$.
This equation is easily solved to give 
$h(t)=h^{\rm FP}_{\rm s}(t)$ where
\be
 h^{\rm FP}_{\rm s}(t)=\frac{J_x(t)-cJ_y(t)}{1-c}. \label{fp}
\ee
Since we must be in the symmetric phase,
the condition $0\le c\le 1$ is imposed on 
the constant parameter $c$.
We note that the Bogoliubov angle $\Theta$ becomes independent of time 
if we impose the fixed-point condition.

Next, we treat the broken phase.
Without losing generality, we can set $J_x\ge J_y$.
The ground-state spins align in the direction denoted by 
the angle variables $\theta$ and $\varphi$.
We consider the rotation in spin space
\be
 &&\hat{U}(\theta,\varphi)
 =\exp\left(-i\theta\hat{\bm{S}}\cdot\bm{n}_0\right), \\
 && \bm{n}_0 = (-\sin\varphi, \cos\varphi, 0),
\ee
so that the ground state is given by the eigenstate of $\hat{S}^z$.
After the rotation, the calculation goes along the same line 
as the case of the symmetric phase.
The only difference is to replace $E_0$, $J_x$, and $h$ by 
\be
 & & \tilde{E}_0=-N\frac{J_x^2+h^2}{2J_x}, \\
 & & \tilde{J}_x=\frac{h^2}{J_x}, \\
 & & \tilde{h}=\sqrt{J_x^2-h^2}+\frac{h^2}{J_x},
\ee
respectively.
The driver Hamiltonian and the fixed-point condition are 
obtained by this replacement.

It should be worth noting that the fixed-point condition can be inferred 
even if we do not know the instantaneous states.
The commutation relation in Eq.~(\ref{commute}) is calculated to give 
\be
 \left[\hat{H}_0(t), \frac{d}{dt}\hat{H}_0(t)\right] 
 &=& 2(J_x\dot{J}_y-J_y\dot{J}_x)\hat{X} \no\\
 &&-2\left((J_x-J_y)\dot{h}-(\dot{J}_x-\dot{J}_y)h\right)\hat{Y},  \no\\
\ee
where $\hat{X}$ and $\hat{Y}$ are time-independent operators
\be
 && \hat{X}=
 \frac{4}{N^2}\left(
 i(\hat{S}^x\hat{S}^y+\hat{S}^y\hat{S}^x)\hat{S}^z
 +(\hat{S}^x)^2-(\hat{S}^y)^2
 \right),  \\
 && \hat{Y}=
 \frac{2i}{N}\left(\hat{S}^x\hat{S}^y+\hat{S}^y\hat{S}^x\right).
\ee
The sufficient condition for this term to vanish is 
that $h$ has the form 
\be
 h(t)=\frac{A+B}{2}J_x(t)+\frac{A-B}{2}J_y(t), \label{fpa}
\ee
where $A$ and $B$ are time-independent constants.
In this case, we have the relation
\be
 \left[\hat{H}_0(t), \frac{d}{dt}\hat{H}_0(t)\right] 
 = 2(J_x\dot{J}_y-J_y\dot{J}_x)\left(\hat{X}-A\hat{Y}\right),
\ee
and the time dependence appears only in the coefficients.
The condition for the coefficient $J_x\dot{J}_y-J_y\dot{J}_x$ to vanish
gives a trivial situation $\hat{H}_0(t)=f(t)\hat{H}_0(t_0)$.
If we consider a state $|\psi\rangle$ such that 
$(\hat{X}-A\hat{Y})|\psi\rangle=0$, the protocol (\ref{fpa})
works as the fixed-point driving.
For example,
in the case of the ground state in the symmetric phase, 
we see from Eq.~(\ref{fp}) that $A=1$.
In the broken phase, we need to perform the unitary rotation 
before calculating the commutation relation.
%Such a task is possible in principle 
%although it is not a simple one.
%Then, the practical way to obtain the critical state 
%will be approaching the state from the symmetric phase.
Thus, we can construct the optimal protocol 
to some extent without knowing the complete solution.
The same consideration can be applied to the model in the previous section.

%%%%%%%%%%%%%%%%%%%%%%%%%%%%%%%%%%%%%%%%%%%%%%%%%%%%%%%%%%%%%%%%%%%%%%%%%%%%
\subsection{Numerical calculation}

%%%%%%%%%%%%%%
\begin{center}
\begin{figure}[t]
\begin{center}
\includegraphics[width=0.9\columnwidth]{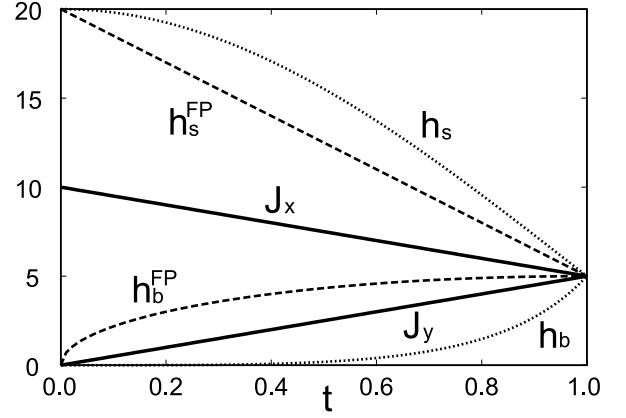}
\end{center}
\caption{Protocols for $J_x(t)$, $J_y(t)$, and $h(t)$ 
to be examined in the numerical analysis.}
\label{fig1}
\end{figure}
\end{center}
%%%%%%%%%%%%

The above analysis is justified only at the thermodynamic limit
and we want to see how the result is sensitive to $N$ and 
is stable against the deviation from the fixed-point condition.

As a simple example, we consider the case with linear time dependence
\be
 & & J_x(t)=10-5t, \\
 & & J_y(t)=5t.
\ee
Starting from the Ising limit $J_y=0$ at $t=0$,
we evolve the state to the isotropic limit $J_x=J_y$ at $t=1$.
Then, we want to bring the system to the critical state 
$J_x=J_y=h$ at $t=1$ stating from a trivial state with 
large or small $h$.
In the symmetric phase, the fixed-point condition is given by 
Eq.~(\ref{fp}) and the initial magnetic field is given by 
$h(0)=10/(1-c)$.
Thus, $c$ is fixed by the initial condition.
In the following, we set $c=0.5$ and 
refer this protocol as $h^{\rm FP}_{s}(t)$.

To compare the result with a different protocol 
with non-fixed-point condition, we examine 
\be
 h_{\rm s}(t) = a+be^{-t^2/2}.
\ee
We set the constants $a$ and $b$ as 
$h_{\rm s}(0)=h_{\rm s}^{\rm FP}(0)$ and 
$h_{\rm s}(1)=h_{\rm s}^{\rm FP}(1)$.

%%%%%%%%%%%%%%
\begin{figure}[t]
\begin{center}
\includegraphics[width=0.9\columnwidth]{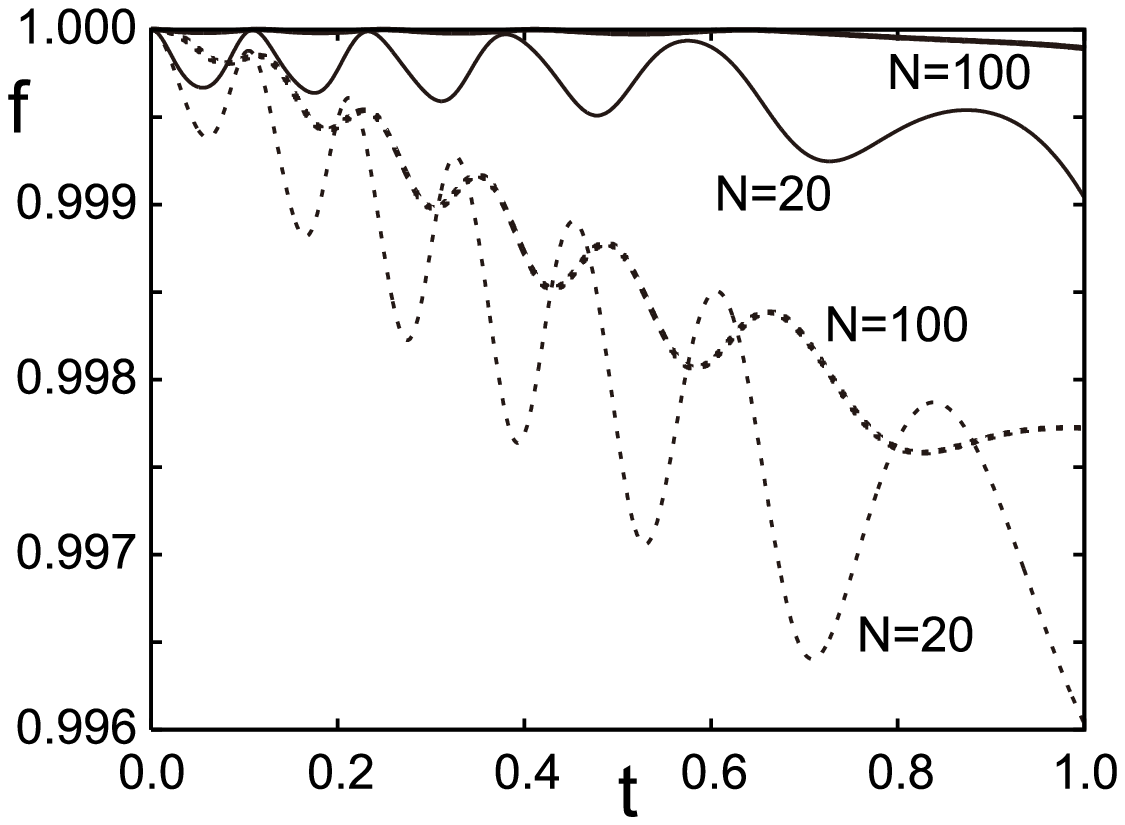}
\includegraphics[width=0.9\columnwidth]{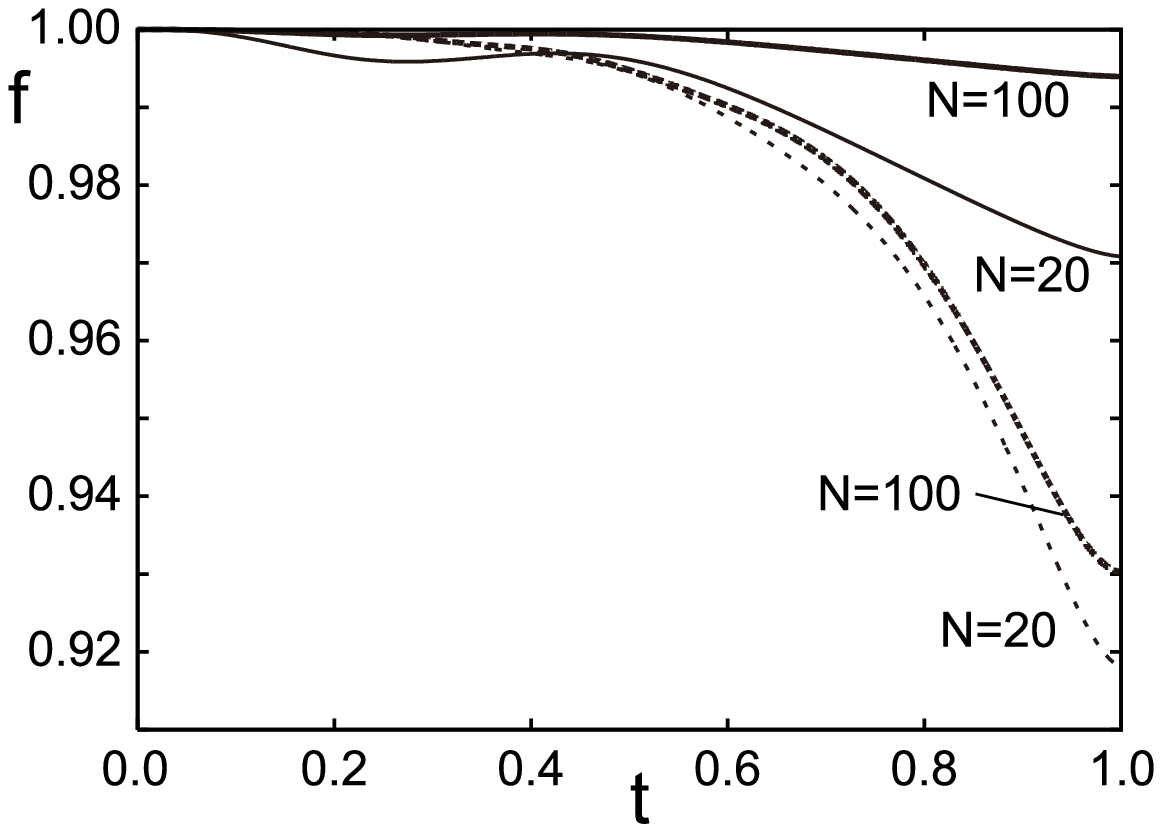}
\caption{Fidelity for the driving of
the ground state in the Lipkin-Meshkov-Glick model.
The solid lines use the fixed-point protocol 
and dotted lines the non-fixed-point one.
The upper figure is for the symmetric phase
and the lower for the broken one.
} 
\label{fig2}
\end{center}
\end{figure}
%%%%%%%%%%%%

We also examine the protocol starting from the broken phase with $h(0)=0$.
From the condition (\ref{fp}) with the replacement 
$J_x\to \tilde{J_x}$ and $h\to \tilde{h}$, 
we can find the fixed-point protocol
\be
 h_{\rm b}^{\rm FP}(t) = \sqrt{J_x(t)J_y(t)}.
\ee
This driving is compared to that of 
\be
 h_{\rm b}(t) = a+be^{t^4}.
\ee
We again use 
$h_{\rm b}(0)=h_{\rm b}^{\rm FP}(0)$ and 
$h_{\rm b}(1)=h_{\rm b}^{\rm FP}(1)$ 
to fix the parameters.
These protocols are summarized in Fig.~\ref{fig1}.

We numerically solve the Schr\"odinger equation to calculate 
the fidelity 
\be
 f = |\langle\psi_{\rm ad}(t)|\psi(t)\rangle|^2, 
\ee
where $|\psi(t)\rangle$ is the exact solution of the equation 
and $|\psi_{\rm ad}(t)\rangle$ is the adiabatic state 
calculated at each time.

The time dependence of the fidelity is summarized 
in Fig.~\ref{fig2}. 
We see that the fixed-point condition works well 
even for the finite system.
The non-fixed-point protocol gives a smaller fidelity which 
confirms our analytic result.
The deviation is not so large and we see that 
the transitionless driving is stable against the deviation.
We confirmed that this observation holds 
for several other possible protocols.
We note that the deviation is large in the broken phase, 
which is because the ground state sensitively depends on time.

%%%%%%%%%%%%%%
\begin{figure}[t]
\begin{center}
\includegraphics[width=0.9\columnwidth]{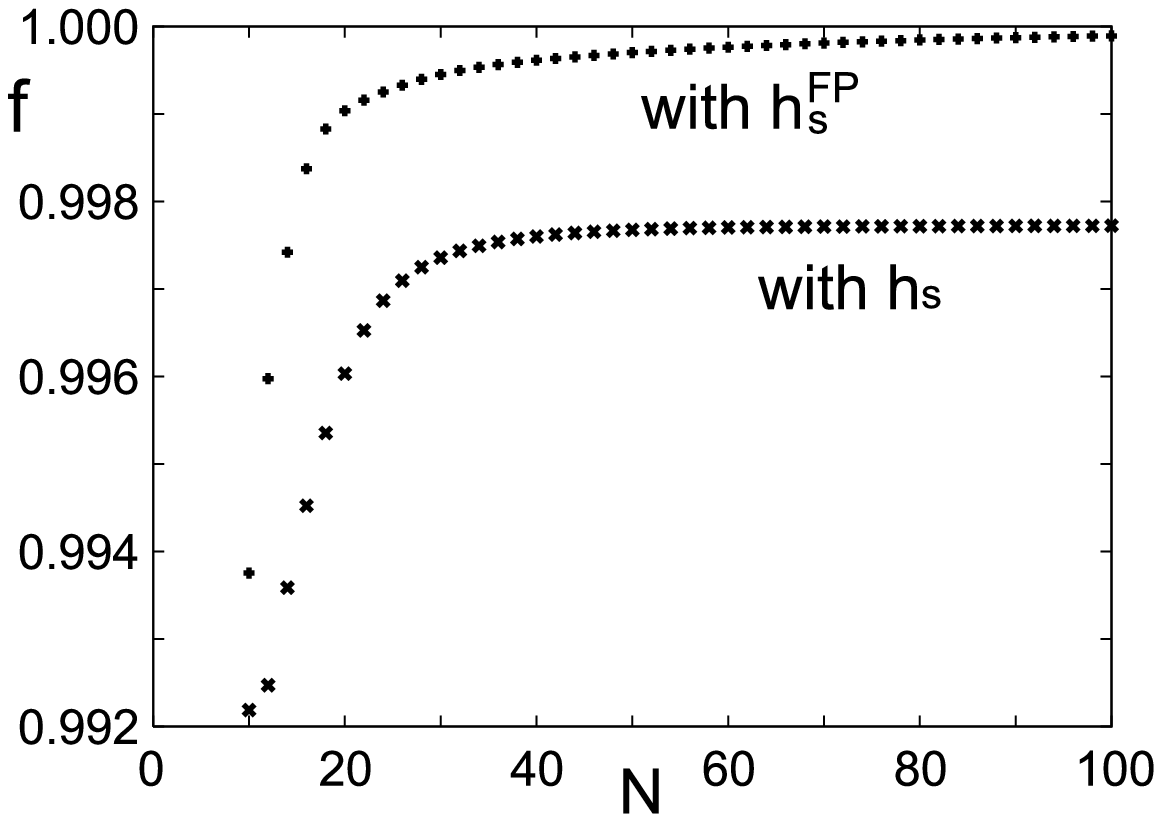}
\includegraphics[width=0.9\columnwidth]{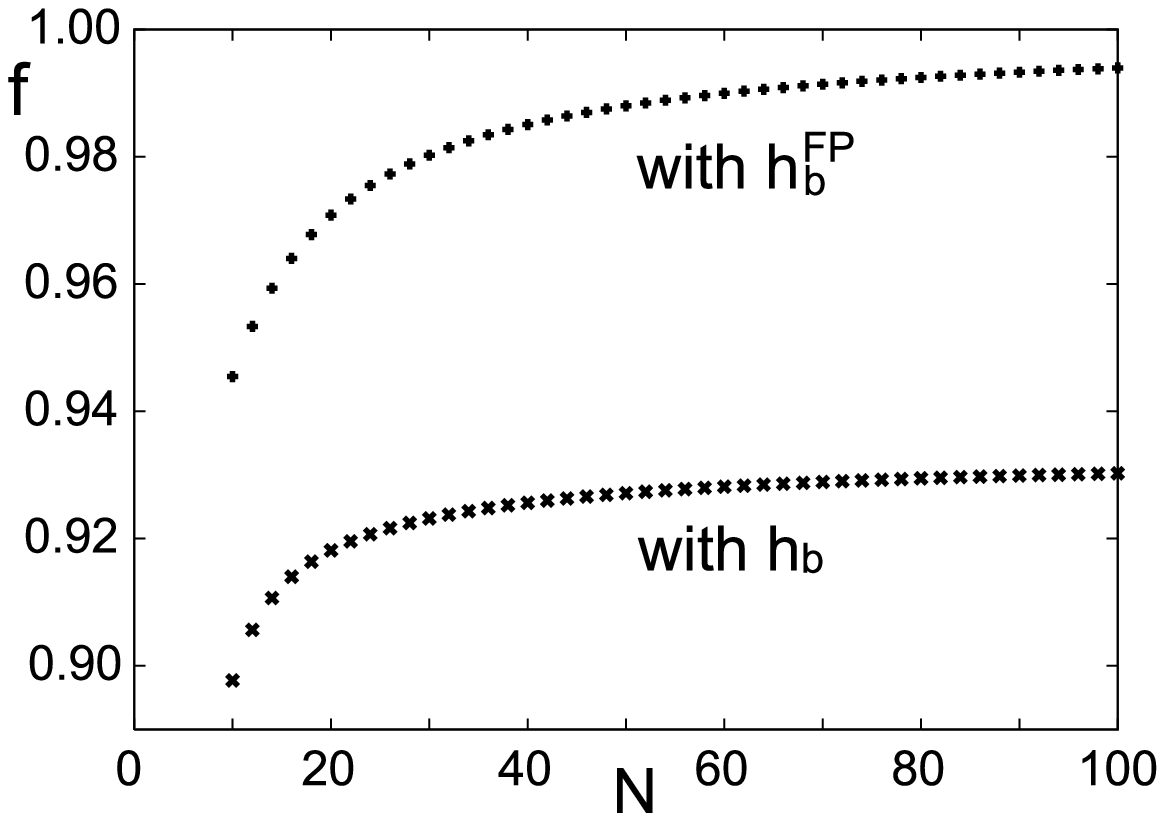}
\caption{The size dependence of the fidelity at the final time $t=1$.
The upper figure is for the symmetric phase
and the lower for the broken one.
}
\label{fig3}
\end{center}
\end{figure}
%%%%%%%%%%%%

The size dependence of the fidelity is shown in Fig.~\ref{fig3}.
The fidelity goes to unity at the thermodynamic limit
in the fixed point protocol and not in the non-fixed-point protocol
as we expect from the analytic result.

%%%%%%%%%%%%%%%%%%%%%%%%%%%%%%%%%%%%%%%%%%%%%%%%%%%%%%%%%%%%%%%%%%%%%%%%%%%%
%%%%%%%%%%%%%%%%%%%%%%%%%%%%%%%%%%%%%%%%%%%%%%%%%%%%%%%%%%%%%%%%%%%%%%%%%%%%
\section{Summary}
\label{sec:summary}

In summary, we have developed the method of the TQD 
and studied possible applications.
We stress that there have not been so many applications of 
this method so far.
In this work we enlarged the applicability of the method to 
various systems.
To establish the usefulness of the present method 
it is necessary to consider more examples 
reflecting realizable experimental situations.

When the method is applied to many-body systems, 
we have mainly three problems.
First, we need to solve the eigenvalue problem 
for the original Hamiltonian which is generally a difficult task.
Second, even if we can find the driving Hamiltonian, 
it has a very complicated form including 
nonlocal and many-body interaction terms.
Third, the driving Hamiltonian diverges 
at the quantum phase transition point where the energy gap goes to zero.
Concerning the first and second problems, 
we have shown that they are eased 
if we consider only a specific state such as the ground state.
In that case, the driving Hamiltonian can be reduced to a simpler one. 
For the third problem, it is possible to avoid 
the divergence by tuning the protocol in a proper way.
Although there is no general prescription, 
we have shown such examples in the one-dimensional $XY$ model
and the Lipkin-Meshkov-Glick model.

The formula of the driving Hamiltonian indicates 
the existence of nontrivial conditions.
Among them, the fixed-point condition is the most interesting one
since the time dependence becomes essentially irrelevant.
For simple systems such as the two-level system, 
this condition only gives a trivial time dependence and is not so useful.
Therefore, the condition is best utilized in many-body systems.
We have found nontrivial examples explicitly in spin systems. 
It can be possible to infer the form of the fixed-point protocol 
by calculating the commutation relation, 
which is simpler than solving the eigenvalue problem.
This will be a guiding factor in developing 
the full dynamics of the many-body systems.

We stress that finding the transitionless Hamiltonian is useful 
not only for direct applications, 
but also for understanding the nature of quantum fluctuations. 
To construct the driving Hamiltonian, 
we need to utilize the full operator space.
For example, in a two-level system, we need 
a magnetic field not in a plane but in the full three-dimensional space.
In some applications such as the quantum annealing,
we usually control Ising spin systems 
by a transverse magnetic field in a single direction. 
The present method clearly indicates that this is not appropriate 
and we should use more different kinds of operators.
We expect that the present method will be a guideline on 
how to construct the optimal Hamiltonian.

Another interesting problem to be studied is 
the robustness of the transitionless Hamiltonian.
It is important to know how much the time evolution is sensitive to 
the control parameters.
This problem was discussed numerically in a specific model in this paper, 
and experimentally in \cite{Betal}.
We need the theoretical ground to understand the result.
This problem will be clarified in future studies.

%%%%%%%%%%%%%%%%%%%%%%%%%%%%%%%%%%%%%%%%%%%%%%%%%%%%%%%%%%%%%%%%%%%%%%%%%%%%
\section*{Acknowledgments}

The author is grateful to T. Obuchi for useful comments. 

%%%%%%%%%%%%%%%%%%%%%%%%%%%%%%%%%%%%%%%%%%%%%%%%%%%%%%%%%%%%%%%%%%%%%%%%%%%%
%%%%%%%%%%%%%%%%%%%%%%%%%%%%%%%%%%%%%%%%%%%%%%%%%%%%%%%%%%%%%%%%%%%%%%%%%%%%

\end{document}